\documentstyle[12pt]{article}
\begin {document}

\begin{flushright} OITS-587\\ 
December 1995\\
\end{flushright}
 
\vskip.5cm
\begin{center} {\Large {\bf Scaling Properties of Hadron Production in the}}
\vskip .1cm {\Large {\bf Ising Model for Quark-Hadron Phase Transition}} 
\vskip .75cm
  {\bf  Zhen CAO, Yi GAO, and Rudolph C. HWA}
\vskip.5cm
 {Institute of Theoretical Science and Department of Physics\\  University of Oregon,
Eugene, OR 97403}
\end{center}

\vspace{.3cm}

\begin{abstract}

Earlier study of quark-hadron phase transition in the Ginzberg-Landau theory is
reexamined in the Ising model, so that spatial fluctuations during the transition can
be taken into account.  Although the dimension of the physical system is $2$, as will
be argued, both $d=2$ and $d=4$ Ising systems are studied, the latter being
theoretically closer to the Ginzberg-Landau theory.  The normalized factorial
moments $F_q$ are used to quantify multiplicity fluctuations, and the scaling
exponent $\nu$ is used to characterize the scaling properties.  It is found by
simulation on the Ising lattice that $\nu$ becomes a function of the temperature $T$
near $T_c$.  The average value of
$\nu$ over a range of $T<T_c$ agrees with the value of $1.3$ derived analytically
from the Ginzberg-Landau theory.  Thus the implications of the mean-field theory
are not invalidated by either the introduction of spatial fluctuations or the restriction
to a 2D system.   
\end{abstract}

\section{Introduction}

It has recently been found in theoretical investigations that in quark-hadron phase
transition (PT) there are hadronic observables that exhibit scaling behavior
\cite{hn}-\cite{h2}.  It is based on the Ginzburg-Landau (GL) description of
second-order PT.  Subsequently, the consideration has been extended to first-order
PT
\cite{h1}-\cite{nz}.  Besides the unique features of the QCD dynamics, what
distinguishes the problem from the standard critical phenomena such as
ferromagnetism is the lack of control of the temperature in a heavy-ion collision and
the availability of only the hadrons as observables.  Together with  the complications
in the nonperturbative regime of QCD, these difficulties preclude at this stage any
observable predictions that follow from first principles.   For that reason the
Ginzburg-Landau theory has been used as the framework for quark-hadron PT, in
which quantitative predictions on the hadronic observables can be derived and
compared with experiment
\cite{hq,hm}.  However, it is a mean-field theory that is accurate only in high
dimensions.  In the Ising model the upper critical dimension is four, above which the
critical exponents determined in the theory is exact.  The physical problem in
heavy-ion collision, however, has dimension $d = 2$, as we shall argue below.  It is
therefore important to consider the Ising model in $d = 2$ for quark-hadron PT, and
compare its scaling behavior, if existent, with that derived in the GL theory.  It is also
useful to consider the $d= 4$ case so that the scaling result can be checked against
the mean-field result.  In that way one can gain some insight on whatever
discrepancy there may exist between the GL and $d=2$ Ising results.  Those are the
problems that we shall investigate in this paper.

The use of Ising model as a solvable problem to study particle production has been
done by many authors in the past several years \cite{jw}-\cite{dsp}.  The focus has
been on the issue of intermittency \cite{bp}, i.e., power-law behavior of the
normalized factorial moments as functions of the bin size.  Differences in their
conclusions appear to depend largely on how a particle is defined in terms of the
lattice spins and how the limit of large $N$ (number of lattice sites) is taken.  In
those papers that are specifically concerned with the $d=1$ Ising model, no PT is
contemplated.  In the GL formalism it has been found analytically that there is no
intermittency
\cite{hn}; nevertheless, there is a different scaling behavior, called $F$-scaling, for
which a numerical scaling exponent $\nu$ has been determined.  We want to check
the value of
$\nu$ in the Ising model.

One of the shortcomings of the GL approach is that the phase transition described by
it is spatially too smooth.  In a heavy-ion collision one expects hadrons to be formed
one at a time or in clusters at the surface of a quark-gluon plasma (if created) with
large spatial fluctuations. The Ising model is well suited to improve on that aspect of
the problem.  We shall consider cells as basic units consisting of several lattice sites,
and vary the size of bins that contain certain multiples of the basic cells.  In that way
the model problem simulates the real situation in a much better approximation.  The
fluctuations that we study cannot be analyzed analytically, so computer simulation
will be essential in carrying out the investigation.  Our study in $d=4$ can help to
elucidate the question whether the difference between the GL and $d=2$ Ising
results is due to the specific dimension chosen or to the modified definition of
hadrons in the Ising system.

\section{Hadronization on the Ising lattice}

The conventional picture of a heavy-ion collision that creates a quark-gluon plasma
is an expanding cylinder of quarks and gluons with temperature $T$ high in the
interior, lower on the surface.  If the plasma undergoes a second-order PT to
hadrons, then there is no mixed phase, and the surface of the cylinder is where the
PT takes place at
$T \leq T_c$, the critical temperature.  After the hadrons are formed, they leave the
plasma and are decoupled from the system under consideration. Thus focusing on the
cylinder surface only, we are dealing with a two-dimensional system of quarks and
gluons, which has patches that hadronize but are immediately replaced by the same
quark system underneath after the hadrons are removed.  The hadronization patches
fluctuate in location and in size, as time evolves.  What we shall analyze is the spatial
fluctuation at any given time.  If time is integrated over, which is equivalent to
collecting all the emitted hadrons from the beginning to the end of the hadronization
process, then the fluctuations we seek would be smoothened out by the overlap of
hadrons produced at different times.  Thus in order to see the fluctuations in the
experimental data  (apart from the complications arising from the possibility of
hadron gas, which is a totally different issue, not to be considered here),  it is
necessary to make $p_T$ cuts.  That is based on the general feature that the average
$p_T$ decreases with increasing time due to the decrease of transverse pressure, as
the longitudinal and transverse expansion progresses.  So selecting a narrow range of
$p_T$ has the effect of selecting a narrow time range of hadronization, during which
the spatial fluctuations can be frozen in the data \cite{hpq}.

The above discussion forms the basis of our consideration of a two-dimensional
system for the second-order quark-hadron PT.  We unfold a finite section of the
cylinder at midrapidity and map it onto a square lattice.  To treat the PT problem on
this lattice in the framework of the Ising model requires a careful consideration of
what constitutes a hadron.  Various definitions have been used in Refs.
\cite{jw}-\cite{dsp}; the most common one associates spin up to a particle at a site. 
That  may be satisfactory for particle production without PT, but  is unsatisfactory 
for our purpose because for $T > T_c$ there should be no hadrons produced except
for those produced due to fluctuating effects.   Our definition is the result of a long
chain of reasoning, which goes as follows.

Our starting point is the Ginzburg-Landau (GL) theory of PT considered in Ref.
\cite{hn}.  One may question the validity of that starting point for hadron
production.  The support for that position is to be found in the production of photons
at the threshold of lasing, which is known to be a second-order PT describable by the
GL theory \cite{gh,dgs}.  Indeed an experiment in quantum optics has accurately
verified the prediction in \cite{hn} of the scaling exponent
$\nu = 1.304$ \cite{yea}.  It is reasonable then, in the absence of a workable theory
to describe the fluctuations of hadronic observables, to replace photons by hadrons
in the GL theory.  In the laser problem the rate of photon emission is
 \begin{eqnarray}
\rho (t) = \left| \phi (t) \right| ^2\quad ,
\label{2.1}
  \end{eqnarray} where $ \phi (t)$ is the complex eigenvalue of the annihilation
operator that specifies the coherent state $|\phi\rangle$.  The probability of
detecting $n$ photons in time $\tau$ with detector efficiency $\eta$ is then
   \begin{eqnarray} P_n(\tau) = \int d\phi d\phi^{\ast} P^0_n\left[ \eta \int^{\tau}_0
\rho(t) dt 
\right] B(\phi)
\label{2.2}
   \end{eqnarray}  where $P^0_n\left[\eta \overline{n} (\tau) \right]$ is the Poisson
distribution corresponding to the multiplicities in a pure coherent state
$|\phi\rangle$ .  The integration in (\ref{2.2}) is weighted by the probability
$B(\phi)$ that the system is in the state $|\phi\rangle$.  That probability is given by
a Boltzmann factor, whose exponent for a homogeneous single-mode laser has exactly
the same form as the GL free energy for second-order PT \cite{gh}-\cite{yea}.

We now adapt this formalism for hadron production with the identification
\cite{h2}
   \begin{eqnarray}
\rho (z) = \left| \phi (z) \right| ^2 \label{2.3}
  \end{eqnarray} as the hadron density at the spatial coordinate $z$ defined in a
$2D$ space.  The Poisson distribution is explicitly 
  \begin{eqnarray} P_n^0= {1 \over n!} \left( \displaystyle \int_A dz\, \rho(z)
\right)^n \mbox{exp} \left[ -\displaystyle \int_A dz\, \rho(z) \right]\quad,
\label{2.4} 
     \end{eqnarray} and the hadron distribution as a result of PT is \cite{hn}
  \begin{eqnarray}  P_n = Z^{-1} \displaystyle \int {\cal D} \phi \, P_n^0\left[ \phi
\right]\, e^{-F \left[ \phi \right]}\quad , \label{2.5} 
     \end{eqnarray}  where $Z = \displaystyle \int {\cal D} \phi \, e^{-F
\left[ \phi \right]}$, and $F[\phi]$ is the GL free energy
 	 \begin{eqnarray} F[\phi] = \displaystyle \int_A dz
\left[ a\left| \phi(z) \right|^2 + b\left| \phi(z)
 \right|^4 + c \left| \partial \phi/ \partial z \right|^2
\right] \quad .\label{2.6}
     \end{eqnarray} Although the gradient term in (\ref{2.6}) is explicitly taken into
account in
\cite{hp}, it is generally not considered in \cite{h2}-\cite{nz}.  The integrations over
$z$ in (\ref{2.4}) and (\ref{2.6}) are over the area of a bin of size $\delta^2$.  Thus
the fluctuations examined are variations from bin to bin without further details
within each bin.  We want to improve on that by use of the Ising model.

In the GL theory of ferromagnetism \cite{bea} the free energy is as given in
(\ref{2.6}) with the order parameter $\phi (z)$ being identified as the local mean
magnetization field, which in the Ising model is given by
   \begin{eqnarray} 
\phi \left( z \right) = A^{-1/2}_{\epsilon} \sum_{j\in A_{\epsilon}} s_j \quad ,
\label{2.7} 
     \end{eqnarray}  where $s_j$ is the spin $\pm\,1$ at site $j$.  The sum in
(\ref{2.7}) is extended over an area $A_{\epsilon}$ (surrounding $z$) smaller than
the resolution of the detector, which in our case is the bin size $\delta^2$, but large
enough to contain many sites.  Hereafter, we shall call that area a {\it cell}, having
the size
$A_\epsilon=\epsilon^2$, where $\epsilon$ will be specified below.  Thus in
connecting quark-hadron PT with the Ising model, using the laser problem and the
GL theory as intermediaries, we should identify the hadron density (\ref{2.3}) with
the absolute square of (\ref{2.7}) so that the hadron multiplicity in the {\it i}th cell is 
 \begin{eqnarray}  n_i =  \lambda \left| \sum_{j \in A_{\epsilon}(i)} s_j\right|^2\, 
\theta (
\sum_j s_j) \quad ,
\label{2.8} 
     \end{eqnarray}  where $\lambda$ is a scale factor that relates the quark density
of the plasma at
$T_c$ to the lattice site density in the Ising model.  
$\lambda$ relates a physical space to a mathematical space, and there is no way to
determine {\it a priori} what that scale factor is.  Any observable consequences of
our calculation should, of course, be independent of $\lambda$.  A Heavyside
function has been appended in (\ref{2.8}) for reasons to be explained below.

The Hamiltonian of the Ising model without external field is 
 \begin{eqnarray}  H = -J\sum_{\left< ij \right>} s_i s_j
\quad,  \label{2.9} 
     \end{eqnarray}  where the sum is over the nearest neighbors.  In the simulation
of the spin configurations the net magnetization $m_L = \sum_{i
\in L^2} s_i$ of the whole lattice of size $L^2$ may be either positive or negative.   If
$m_L < 0$, we reverse all spins, replacing $s_i$  by $-s_i$ at every site in the lattice,
so that we always have $m_L \geq 0$, essentially by definition.  We then define
hadron multiplicity in a cell by (\ref{2.8}), in which the $\theta$-function puts $n_i =
0$, when the net spin in cell $i$ is not positive definite.  Thus at $T > T_c$ the
disordered state tends to have $n_i$ near zero; even if $n_i \not= 0$ for certain cells
due to fluctuations, there should be nearly as many cells in a bin that have no
hadrons.  At $T < T_c$ the ordered state has mostly nonempty cells.  These are the
characteristics that are physically reasonable, and should be incorporated in the
description of the heavy-ion collisions where hadrons are produced from the
cylinder surface in patches of fluctuating positions and sizes.  It is an aspect of the
hadronization process that cannot be addressed in the GL theory.  
\section{Scaling behavior in the Ising model in $d = 2$}

Monte Carlo simulation of the Ising model by use of the Metropolis algorithm suffers
greatly from the phenomenon of critical slowing down, when $T$ is near
$T_c$
\cite{bea}, i.e. the autocorrelation time in generating new configurations becomes
very long.  Previous numerical work on the study of intermittency in the Ising model
\cite{jw,bfs} must have encountered that difficulty.  We avoid the problem entirely
by adopting the Wolff algorithm \cite{wlf}, with which there is no dependence on the
number of steps required to remove autocorrelation.

We work  with a 2D lattice of size $L = 288$ and consider cells of  size
$\epsilon = 4$.  We use $10^4$ sweeps to set up an initial configuration, and then
generate $10^4$ configurations to calculate event averages.  The average hadron
multiplicity
$\left< n\right>_{\epsilon}$
 in a cell in units of $\lambda$ is shown as a function of $T$ in Fig.\ 1.  The sharp
drop occurs at $T = 2.3$\, (in units of $J/k_B$), a value slightly higher than the
analytical value of $T_c$, 2.27, for infinite lattice \cite{hsm}.  Our determination of
the precise value of $T_c$ will be given below according to a well-defined
procedure.  In Fig.\ 1, $\left< n\right>_{\epsilon}/\lambda$ approaches 256 as $T
\rightarrow 0$, as it should, since when all spins are aligned, the total-spin squared
in a cell is $(4 \times 4)^2$.  In the large $T$ limit
$\left< n\right>_{\epsilon}$ is not zero because not every cell has equal numbers of
spin-up and -down.  However, the nonvanishing of
$\left< n\right>_{\epsilon}$ is not important for our result, for we shall examine
scaling behaviors only in the neighborhood of $T_c$.

The observable whose scaling properties are to be studied is the normalized factorial
moments \cite{bp}
\begin{eqnarray}  F_q = \left< {\left< n_{\delta} \left(n_{\delta} - 1 \right) \cdots
\left(n_{\delta} - q +1
\right) \right>_h
\over  \left< n_{\delta}\right>^q_h} 
\right>_v \quad .
\label{3.1} 
     \end{eqnarray}  where $n_{\delta}$ is the hadron multiplicity in a bin of size
$\delta ^2$:
		\begin{eqnarray}  n_{\delta} = \sum^{N_\delta}_{i = 1} n_i \quad , \label{3.2} 
     \end{eqnarray}
 $n_i$ being the cell multiplicity given in (\ref{2.8}), the sum being over all 
$N_\delta$ cells in a bin, where $N_\delta = (\delta/\epsilon)^2$.  In (\ref{3.1})
$\left< \cdots
\right>_h$ means (horizontal) average over all bins in a configuration, and $\left<
\cdots
\right>_v$ means (vertical) average over all configurations.   The number of bins
$M$ in the lattice is
$M=(L/\delta)^2=(L/\epsilon)^2/N_\delta$. Intermittency refers to the scaling in the
bin size or in
$M$
  \begin{eqnarray}  F_q (M) \propto M^{\varphi _q}  \label{3.3} 
     \end{eqnarray}  over a range of $M$.  More precisely, we shall  refer to it as
$M$-scaling to be distinguished from $F$-scaling, which is the power-law behavior
   \begin{eqnarray}  F_q \propto F_2^{\, \beta_q}, \label{3.4} 
     \end{eqnarray}  when $M$ is varied.  Clearly, the validity of (\ref{3.3})
guarantees (\ref{3.4}), but (\ref{3.4}) may be valid even if (\ref{3.3}) is not.  The
latter situation is, in fact, the case in the GL theory \cite{hn}.  Phenomenologically,
there is support for
$F$-scaling, when the evidence for $M$-scaling is weak, from data that have no
relevance to quark-hadron PT \cite{owo}.

To determine the $F_q$ moments we generate a spin configuration at a given
$T$ by the Wolff algorithm and calculate the cell multiplicity $n_i$ for a wide range
of
$\lambda$ for all cells in the lattice, which is partitioned into $\left(
L/\delta\right)^2$ bins.  Since for any given $\lambda$ the bin multiplicity
$n_{\delta}$ need not be an integer, we take the largest integer less than
$n_{\delta}$ and use it in (\ref{3.1}) for each bin.   We then perform the  horizontal
averages in (\ref{3.1}) over all bins in the lattice, followed by averaging over
$10^4$ configurations at the same $T$.  The dependence of $F_q(M)$ on $\lambda$ is
shown in Fig.\ 2 for a number of representative values of $q$ and $M$.  As expected,
there is nearly no dependence on $\lambda$ when $\lambda$ is large in the interval
$(0,1)$, since when $n_{\delta}$ is large the factorial product $n_{\delta}
(n_{\delta}-1) \cdots (n_{\delta}-q+1)$ is approximately equal to $n_{\delta}^q$ and
the $\lambda$ factors are canceled in the ratio.  But when
$\lambda$ is very small, then the effect on $F_q$ is significant.

Since the critical point is independent of how we map the quark density to the lattice
density, we first set $\lambda = 1$ and determine the critical temperature
$T_c$.   In Fig.\ 3 we show the $M$ dependence of $F_q(M)$ for $\lambda = 1$ at
three values of $T$: 2.20, 2.315, and 2.40, always in units of $J/k_B$ hereafter.
  It is evident that $M$ scaling is good for $T = 2.315$, but not for the other two
temperatures. The quality of straightness in the log-log plots can be made more
transparent by comparing just the $q = 6$ moments in the same figure, as shown in
Fig.\ 4.  Since at $T_c$ clusters of hadronic patches of all sizes are formed, we expect
$F_q$ to exhibit $M$ scaling at $T_c$  as a characteristics of criticality.  Thus we have
a criterion for the determination of $T_c$: $F_q(M)$ should show strict $M$ scaling at
$T_c$.  In Fig.\ 5 we show the
$\chi^2$ of straightline fits of all $F_q$ $(2
\leq q \leq 6)$ for a range of $T$ in the neighborhood of small $\chi^2$ and find that
$\chi^2$ is lowest at
$T = 2.315$, which is therefore the $T_c$ in this problem.

Having determined $T_c$, we now investigate the $\lambda$ dependence of $M$
scaling.  In Fig.\ 6 we show at $T_c$ the $M$-scaling behaviors for six values of
$\lambda$:  1, 0.1, 0.05, 0.03, 0.02, and 0.01.  For each value of $q$ the set of curves
appear lower for lower values of $\lambda$.  Clearly, $M$ scaling is still good up to
$M \simeq$ 100 even for the lowest $\lambda$, but for $M > 100, \ F_q$  bends over
and even decreases at high $M$ and small $\lambda$.  This behavior is reasonable,
because at small $\lambda$, the cell multiplicities $n_i$ are small, so the bin
multiplicities $n_{\delta}$  are also small at high $M$; thus the factorial product in
(\ref{3.1}) vanishes for many cells at high $q$, and $F_q$  receives nonvanishing
contribution only from the rare cells and events that have large fluctuations.  This
phenomenon is not an artifact of the Ising model, but is the realistic situation in the
data on multiparticle production (see, for example, the Workshop proceedings of
\cite{chs,hos}).  For this reason our use of the Ising model in the way we define
hadron multiplicities constitutes a concrete improvement over the previous work on
the Ising model on the one hand, and over the GL approach to quark-hadron PT on
the other.

From Fig.\ 6 we can determine the slopes $\varphi _q$ from the regions where
$M$ scaling is good.  Even for $\lambda$ = 0.01, that can be done between $M$ = 10
and 100.  The results for $\lambda$ = 1.0 and 0.01 are shown in Fig.\ 7.  Evidently,
the dependence of $\varphi _q$ on $q$ is remarkably linear, which we parameterize
as
    \begin{eqnarray} 
\varphi _q = \zeta(q - 1) \quad . \label{14} 
     \end{eqnarray} The slopes determined from Fig.\ 7 are 0.135 and 0.121,
respectively, which can be summarized as $\zeta = 0.128 \pm 0.007$.  This result
should be compared to that of the analytical work of Satz on the Ising model
\cite{hs}, where $\zeta = 1/8$ for $d = 2$.  The agreement is excellent despite the
fact that no reference to hadrons is made in \cite{hs}.  This gives us confidence in our
procedure of relating quark-hadron PT to the Ising model.

The above result is for $T=T_c$ only.  As mentioned earlier, there is no $M$ scaling
at  $T\not=T_c$.  We now show that $F$ scaling as defined in (\ref{3.4}) is
nevertheless good for a small range of $T\leq T_c$.  Furthermore, the slope
$\beta _q$ satisfies the formula 
   \begin{eqnarray} 
\beta_q = (q - 1) ^{\nu}  \label{15} 
     \end{eqnarray}  found in \cite{hn}.  But first, at $T_c$ the validity of (\ref{3.3})
and (\ref{14}) implies the validity of (\ref{3.4}) with 
   \begin{eqnarray} 
\beta_q = \varphi _q/\varphi _2 = q - 1\quad . \label{16} 
     \end{eqnarray} Thus at  $T=T_c$, we have $\nu = 1$ exactly to the extent that
(\ref{14}) is exact.  Of course, (\ref{14}) is only a fitted result that is very good, but
not exact.

A direct evaluation of $F$ scaling can be accomplished by examining $F_q$ vs
$F_2$ in log-log plots.  In Fig.\ 8 we show those plots for three values of $T$ at
$\lambda = 1$: (a) $T = 2.20$, (b) $T=T_c = 2.315$, and (c) $T = 2.40$.  It is clear that
$F$ scaling is good for the first two cases, but not for the third.  Hadronization at $T >
T_c$ is only due to fluctuations, and we shall not consider that region any more
below.  On the other hand, PT can take place at $T$ in a small region below
$T_c$, so our $F$-scaling analysis takes into account a wider range of hadrons
produced in a PT, not just at $T_c$.

The values of the slopes $\beta_q$ in Fig.\ 8(a) and (b) are shown by square symbols
in Fig.\ 9 as functions of
$q-1$ in the log-log plot .  They are well fitted by straight lines, thus verifying the
validity of (\ref{15}).  The exponents $\nu$ are found to be 1.068 for
$T = 2.315$, and 1.565 for $T = 2.20$.  These are the results for $\lambda = 1$.

To see the $\lambda$ dependence, we show in Fig.\ 10 the $F$-scaling plots for four
values of $T$ at and below $T_c$ and for various values of $\lambda$.  The range of
$\lambda$ in which $F$-scaling is manifest shrinks as $T$ is moved farther away
from  $T_c$.  For those values of $T$ and $\lambda$ where straight-line fits can
reasonably be made in those figures, the slope parameters
$\beta _q$ are determined.  They all satisfy (\ref{15})  very well.  The case for
$\lambda = 0.05$ is shown by diamond symbols in Fig.\ 9.  Thus
$F$  scaling, when it exists, can be represented by a value of the scaling exponent
$\nu$, which depends on
$T$ and $\lambda$.  In Fig.\ 11 is shown $\nu (T)$  for various $\lambda$.  When no
value of $\nu$ is given, it means that $F$-scaling is not good enough to provide
unambiguous extraction of $\beta _q$.  For $\lambda = 1$ the range of $T$ where
$\nu$  can be determined is largest.  Fig.\ 11 shows that $\nu$ increases from 1.07
at $T_c$ to 1.67 at $T = 2.1$.  But for  $\lambda = 0.02$ only at $T_c$ can $\nu$ be
determined to be 1.02.  The result at $T_c$, i.e., $\nu (T_c) = 1.02 - 1.07$, differs
slightly from the earlier result of $\nu = 1$ that follows from (\ref{16}).  They are
close enough to be compatible, considering the fact that both procedures involve
fitting of approximate scaling curves in different variables.

The physical situation corresponds to low values of $\lambda$, since the hadron
multiplicities in a cell should not be too large.  Taking only $\lambda = 0.02$ and
$0.05$, we show the corresponding values of $\nu$ in Fig.\ 12.  The variation is
between $\nu = 1.02$ and $1.52$ over a narrow range of $T$ below $T_c$.  It
straddles very well the value $\nu = 1.304$ (dashed line) determined in the GL
theory \cite{hn}.  We recall that in the GL calculation there is no dependence of
$\nu$ on the GL parameters $a(T)$ and $b$ (so long as $T \leq T_c$) and on the
dimension $d$.  It is an average result that is situated in the middle of the range of
$\nu(T)$ that we have now determined with the spatial fluctuations taken into
account.  The two results therefore give mutual support to each other.

We have varied the cell size by reducing $\epsilon$ from 4 to 3.  All the results are
found to be virtually unchanged.  Indeed, a sensitive dependence of $\epsilon$
would render our procedure of determining the scaling behavior of particle
production unreliable and physically irrelevant.

Since $T$ is not subject to experimental control in heavy-ion collisions, we have no
way to check the $T$ dependence of $\nu(T)$ empirically.  Detected hadrons, even in
a small $p_T$ interval may be created in a range of
$T$ below $T_c$, so the observation of $\nu = 1.3$ for all hadrons emitted would be
a confirmation of both the GL and Ising descriptions of PT.  Our study here not only
generalizes the PT problem to allow for spatial fluctuations of hadronization patches,
but also demonstrates that smaller $\nu$ means larger fluctuation, since it is known
that the fluctuation is greatest at $T_c$.

\section {Scaling behavior in $d=4$}

Before going into $d=4$, let us first review what we have learned so far.  We have
been trying to relate three physical processes and models:  (a) quark-hadron PT, (b)
GL theory of PT, and (c) Ising model in $d=2$.  From (b) is derived in \cite {hn} the
scaling behavior of
$F_q$, which is the measure appropriate for (a); however, being a mean-field theory
(b) is too smooth to reveal spatial fluctuations of (a).  Since the physical dimension
for the PT is $d=2$, (c) is used to examine the scaling behavior where the
discreteness of the hadron multiplicity in a cell is emphasized.  The result affirms the
$F$-scaling behavior found in (b), but with the scaling exponent
$\nu$ dependent on $T$, which is not a feature of (b).  At $T_c$ the value of $\nu$
agrees with what can be inferred from the analytic treatment of 2D Ising model \cite
{hs}, but when averaged over a small range of $T<T_c$, it agrees with the analytic
result derived on the basis of (b).   Thus the disagreement between the original,
strict results of (b) and (c) (i.e., $\nu=1.3$ and 1.0, respectively) is removed by our
present treatment of hadron formation in the context of (a).  However, there is a
level of understanding still unattained thus far.  In carrying out our calculation we
have used the definition of hadrons on the lattice as described in Sec.\ 2 at the same
time when the 2D Ising model is used to simulate second-order PT.  The question
then arises as to which one of these two aspects of the problem is more responsible
for the result of the previous section, i.e., our way of counting hadrons or the
two-dimensionality of the space.  To answer this question we perform a similar
calculation using the hadron definition of Sec.\ 2, but in the $d=4$ Ising model, which
is known to have critical indices in agreement with the GL theory.  In so doing we
can learn whether the $T$ dependence of $\nu$ is a result of discrete hadron
counting (as would be the case if the behavior persists in $d=4$) or of the
two-dimensionality (if the behavior disappears in $d=4$).

In $d=4$ we must be content with smaller lattice size.  We have used
$L=24$ in each dimension, and considered a cell as being of volume
$\epsilon^4$, with $\epsilon=2$.  Thus a cell has $16$ sites, as in the case of $d=2$
where $\epsilon=4$.  Our bins of lattice volume
$\delta^4$ can contain $N_{\delta}=(\delta/ \epsilon)^4$ cells, which can vary from
$N_{\delta}=1$ to $12^4$ cells.

As before, we use (8) to define the hadron multiplicity on the 4D lattice, except, of
course, that the sum is over $j\in V_{\epsilon}(i)$, all sites of a 4D cell at $i$.  Since a
4D lattice is far from the reality of a quark-gluon system, there is no precise meaning
to the value of
$n_i$ calculated by (8), apart from the physical interpretation that it plays the role
of multiplicity in 4D.  The scale factor $\lambda$ should, nevertheless, be present as
in the 2D case to make the present calculation as close to the previous one as
possible, so
$\lambda$ can be varied over the same range.  As in 2D, we would not want to think
that a cell can contain as many as $256$ particles.

We continue to use (9) as the Hamiltonian, but in 4D, and to employ the Wolff
algorithm for simulation.  Since we have been unable to find in the literature an
analytical determination of
$T_c$ for the 4D Ising model, we have to search for the critical point without any
guidance except by the procedure used in the previous section.  

In Fig.\ 13 we show the average cell multiplicity in units of $\lambda$ as a function
of $T$.  It is the result of averaging over $3\times10^4$ configurations after initial
runs of $3\times10^4$ sweeps.  Evidently, the transition temperature is not as
distinct as in Fig.1 for $d=2$.  To determine $T_c$ precisely we have attempted to
find the temperature at which there is $M$ scaling as in Fig.\ 3(b).  Since $T_c$ is
independent of $\lambda$, we set $\lambda =1$ during the search.  Unfortunately,
we have not been able to find any $T$ at which $F_q$ exhibits strict power-law
behavior in $M$, which is now $M= (L/ \delta)^4 = (L/
\epsilon)^4/N_{\delta}$.  Perhaps it is because in 4D the value of $L$ at $24$ is not
large enough to allow the full manifestation of long-range correlation.  Fig. 14 is an
example of the $M$ dependence of $F_q$.  That figure is actually for
$T=T_c$, the value of which is to be determined later.  At neighboring values of
$T$ the behaviors of $F_q$ vs $M$ are similar, so we omit their presentation.

For any set of $F_q$ at a given $T$ we can plot $F_q$ vs $F_2$.  Fig.\ 15 shows such
a plot for the $F_q$ in Fig.\ 14.  Fitting all such plots by the $F$-scaling formula (13)
we can determine the scaling exponent $\nu$. Fig.\ 15 shows the dependence of
$\nu$ on $T$.  Remarkably there is a dip at $T =6.7$.  Mindful of a similar dip in
Fig.\ 11, we declare the critical point to be at $T_c =6.7$.  In the absence of any other
method that we can apply for a precise determination of $T_c$, the dip in $\nu(T)$
emerges as the only feasible method.  Knowing $T_c$ allows us to focus on the
behavior of the scaling exponent in its neighborhood.

Returning to Fig.\ 14 we can examine the $M$ dependences for various values of
$\lambda$ at $T_c$.  Note that there is essentially no difference between
$\lambda =1.0$ and $0.1$, just as it is in Fig.\ 6 for $d=2$.  But for $\lambda =0.01$
all $F_q$ fall below $1$ for $M>10^4$ and are not plotted in Fig.\ 14.  For the range
of $M$ shown for $\lambda =0.01$, which corresponds to 
$4\leq\delta\leq12$, the dotted lines almost show $M$ scaling.  In the $F$-scaling
plot of Fig.\ 15, there appears to be no dependence on $\lambda$ for $0.01\leq
\lambda\leq1.0$ up to the second highest values for $F_q$.  The last and highest
points for log$ F_2 >0.6$ are for $\delta=\epsilon=2\ (M= 2\times10^4)$ and for
$\lambda= 1.0$ and $0.1$ only.  The large gap from the second highest ones
$(\delta =2\epsilon)$ cannot be reduced because the bin size must jump discretely
from two cells to one in each dimension.  The values of $\nu$ are determined by
straight-line fits of the portion of Fig.\ 15 corresponding to $\delta\geq4$, i.e.,
excluding the highest points.

The result on $\nu (T)$, shown in Fig.\ 16, exhibits some $\lambda$ dependence at
$T_c$.  For the case $\lambda =0.01$, which is perhaps closest to the physical
problem with the lowest $\langle n\rangle$ per cell, we have $\nu \approx 1.3$, in
agreement with the result in GL theory \cite {hn}.  That is gratifying, since this is our
only way of checking that the scaling behaviors of 4D Ising model and of the GL
theory agree.  The fact that $\nu$ is distinctly different from $1.0$ implies that the
scaling exponent is not universal.  As in the 2D case, $F$ scaling is valid only within a
very small range of $T$ around $T_c$, when $\lambda =0.01$.

When $\lambda\geq 0.1$, we can investigate the $T$ dependence of $\nu (T)$.  Fig.\
16 shows that $\nu (T)$ increases as $T$ decreases from $T_c$, exhibiting essentially
the same features as in Fig.\ 11.  This result answers the question we have asked
toward the end of the first paragraph of this section.  The $T$ dependence of $\nu
(T)$ is independent of the dimension $d$.  Since the analytic result from the GL
theory is that $\nu =1.3$ independent of $T$ \cite {hn}, we now see that the $T$
dependence of $\nu (T)$ is a consequence of the spatial fluctuations, which have
been simulated on the lattice, but are lacking in the GL theory, and has little to do
with the dimension of the system.  

\section {Conclusion}

By working with the Ising model we have learned a number of things about
second-order PT that are relevant to the quark-hadron problem.  First of all, because
the physical system of interest is two dimensional, it is the result of the 2D Ising
model that we should concentrate on.  Using the definition of hadrons on the lattice
that we have constructed on the basis that hadron density is related to the square of
the order parameter (or the total magnetization in a cell), we have been able to
introduce spatial fluctuations during the hadronization process, a feature that is
missing in the GL description.  In so doing we are faced with a dilemma that the
result of our simulation resolves.

The dilemma is that the scaling behavior of 2D Ising model is known from analytical
studies
\cite {hs} to imply $\nu =1.0$, whereas the GL theory yields the analytical result of
$\nu =1.3$ \cite {hn}.  The resolution is that when discrete hadron counting is
achieved by lattice simulation, a temperature dependence is introduced into
$\nu(T)$.  At $T=T_c$, $\nu (T_c)$ is indeed approximately $1.0$, but at $T<T_c$,
$\nu(T)$ is larger.  The average is approximately
$1.3$, as shown in Fig.\ 12.  In the GL theory $\nu$ is independent of $T$ so long as
$T<T_c$, so its result $\nu=1.3$ may be regarded as an average over $T$ in
agreement with our simulated result.  In the physical problem of quark-hadron PT in
heavy-ion collisions, the hadronization temperature is not measurable.  The observed
hadron multiplicity fluctuations may well reflect hadronization occurring at a range
of $T<T_c$.  Thus the average $\nu=1.3$ is still the relevant scaling exponent in
phenomenology.

The $T$ dependence of $\nu(T)$ is not a special feature of $d=2$.  By simulating in
$d=4$ we have found essentially the same $T$ dependence.  Moreover, at $T_c$ in
4D the value of
$\nu(T_c)$ is approximately $1.3$ rather than $1.0$, reconfirming the known fact
that the GL theory and the Ising model agree at $d=4$ or higher.  This gives us
confidence in our choice of definition of hadron multiplicity in the Ising model and in
the physical implications of the results obtained by simulation in 2D.  

The basic understanding we have learned is that when the shortcoming of the GL
theory is amended by taking into account the spatial fluctuations of hadronization,
the predicted result of
$\nu=1.3$ is unchanged, except for the realization that it is a mean value after
averaging over a range of $T$ below $T_c$.

It is also important to recognize at a more general level that in our attempt to relate
quark-hadron physics to statistical physics we have found the effectiveness of using
the factorial moments $F_q$, the significance of $F$-scaling, and the amazing validity
of $\beta_q= (q-1)^{\nu}$ in all circumstances, with the consequence that the scaling
exponent $\nu$  has emerged as an effective parameter characteristic of PT, in lieu
of the usual critical indices that rely on $T$ being under experimental control.  It is
therefore suggestive that our method of describing the properties of PT may find
applications in other areas of science where discrete multiplicities are the only
observables at one's disposal.  Such an application has recently been found in
evolutionary biology \cite {chw}.

\vskip.7cm
\centerline{\large \bf Acknowledgment}
\vskip.5cm This work was supported, in part, by the U.S. Department of Energy
under grant No. DE-FG06-91ER40637.

\newpage

\centerline{\large \bf Figure Captions}

\begin {description}

\item [Fig.\ 1]\quad  Average multiplicity in a cell ( in units of $\lambda$) as a
function of temperature for
$d=2$ lattice. 

\item [Fig.\ 2]\quad  Normalized factorial moments as functions of the scales factor
$\lambda$ for
$q=3$ and
$6$ and for three values of $M$.

\item [Fig.\ 3]\quad  $F_q$ vs $M$ at three temperatures.

\item [Fig.\ 4]\quad  Same as in Fig.\ 3 but for $q=6$ only.

\item [Fig.\ 5]\quad  $\chi^2$ of straight-line fits of log$F_q$ vs log$M$ for
$T$ in the neighborhood of
$2.315$.

\item [Fig.\ 6]\quad  $M$-scaling plots of $F_q$ at $T_c$ for various values of
$\lambda$.  In each cluster for a fixed $q$ lower curves have lower $\lambda$.

\item [Fig.\ 7]\quad  Intermitting indices $\varphi_q$ at $T_c$ for $\lambda =1.0$
and $0.01$.

\item [Fig.\ 8]\quad  $F$-scaling plots at three temperatures.

\item [Fig.\ 9]\quad  The slope $\beta_q$ in log-log plot for different $T$ and
$\lambda$.

\item [Fig.\ 10]\quad  $F$-scaling plots for different $T$ and $\lambda$.

\item [Fig.\ 11]\quad Scaling exponent $\nu(T)$ for different $\lambda$.

\item [Fig.\ 12]\quad  $\nu(T)$ at small $\lambda$ compared to the analytical value
$\nu=1.304$ from [1].

\item [Fig.\ 13]\quad  Average multiplicity in a cell ( in units of $\lambda$) as a
function of $T$ for
$d=4$ lattice.

\item [Fig.\ 14]\quad  $M$-scaling plots for different $\lambda$ at  $T_c$ in 4D.

\item [Fig.\ 15]\quad  $F$-scaling plots for different $\lambda$ at $T_c$ in 4D.

\item [Fig.\ 16]\quad $\nu(T)$ near $T_c$ for different $\lambda$ in 4D.    

\end {description}

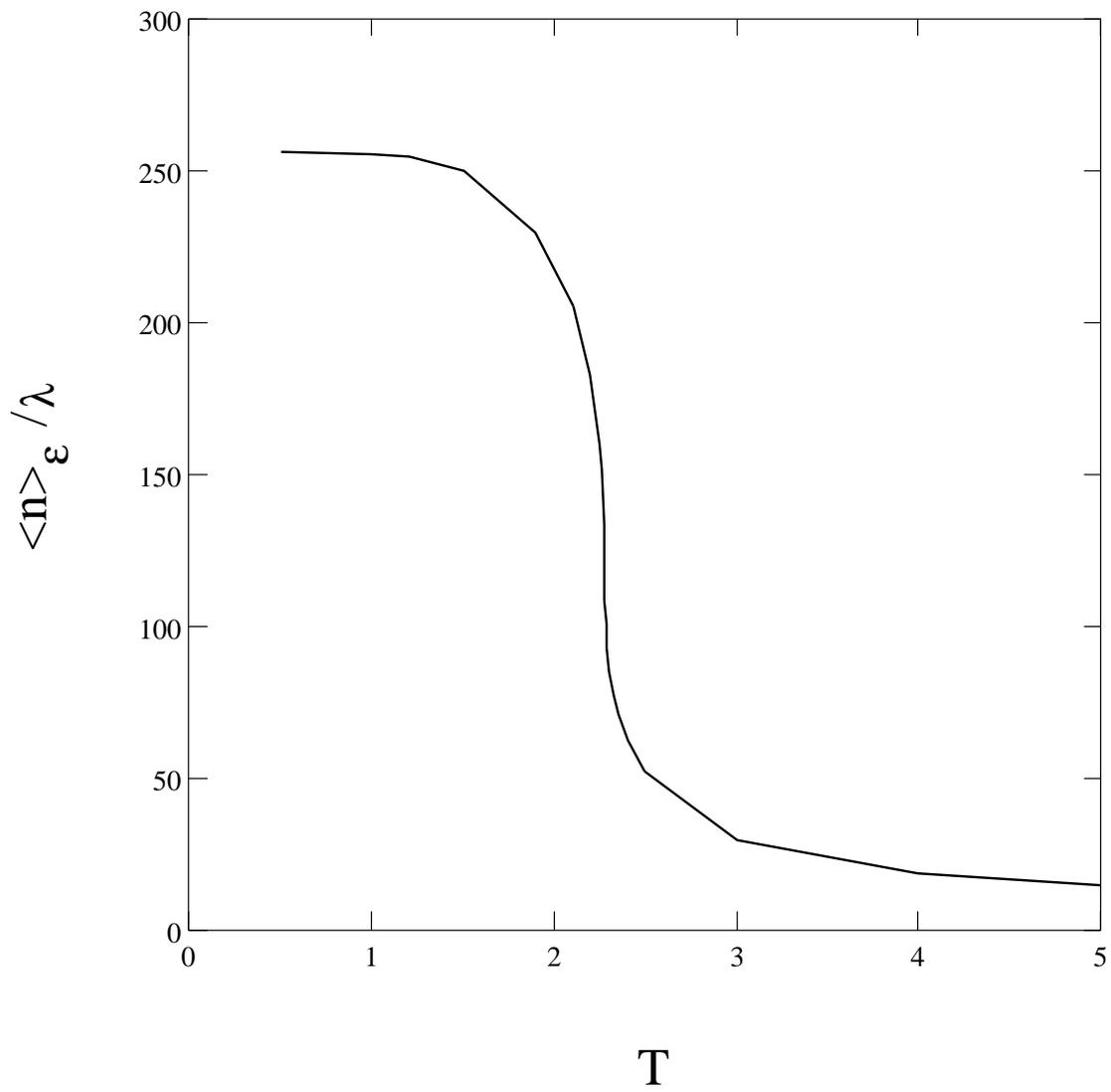
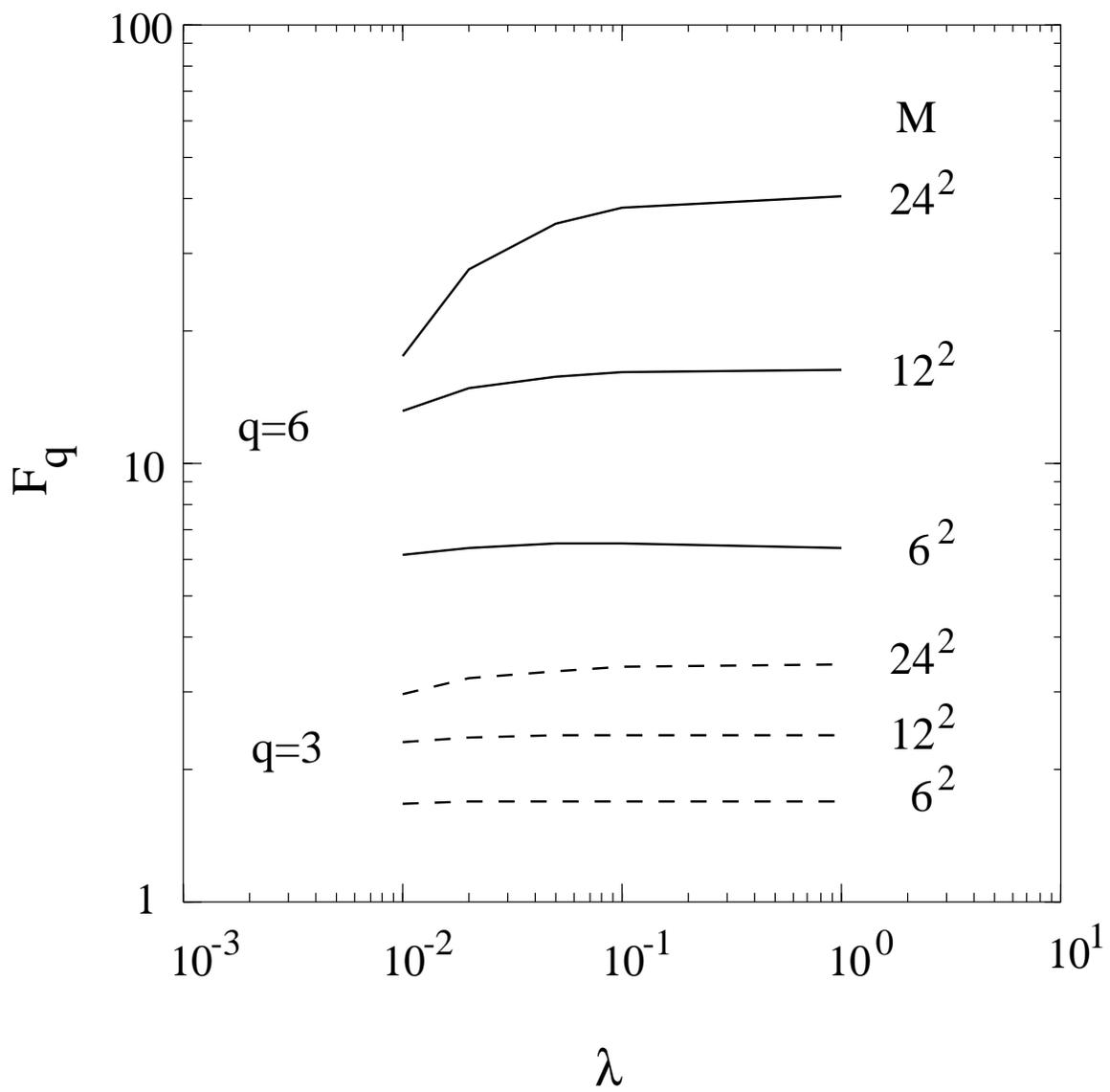
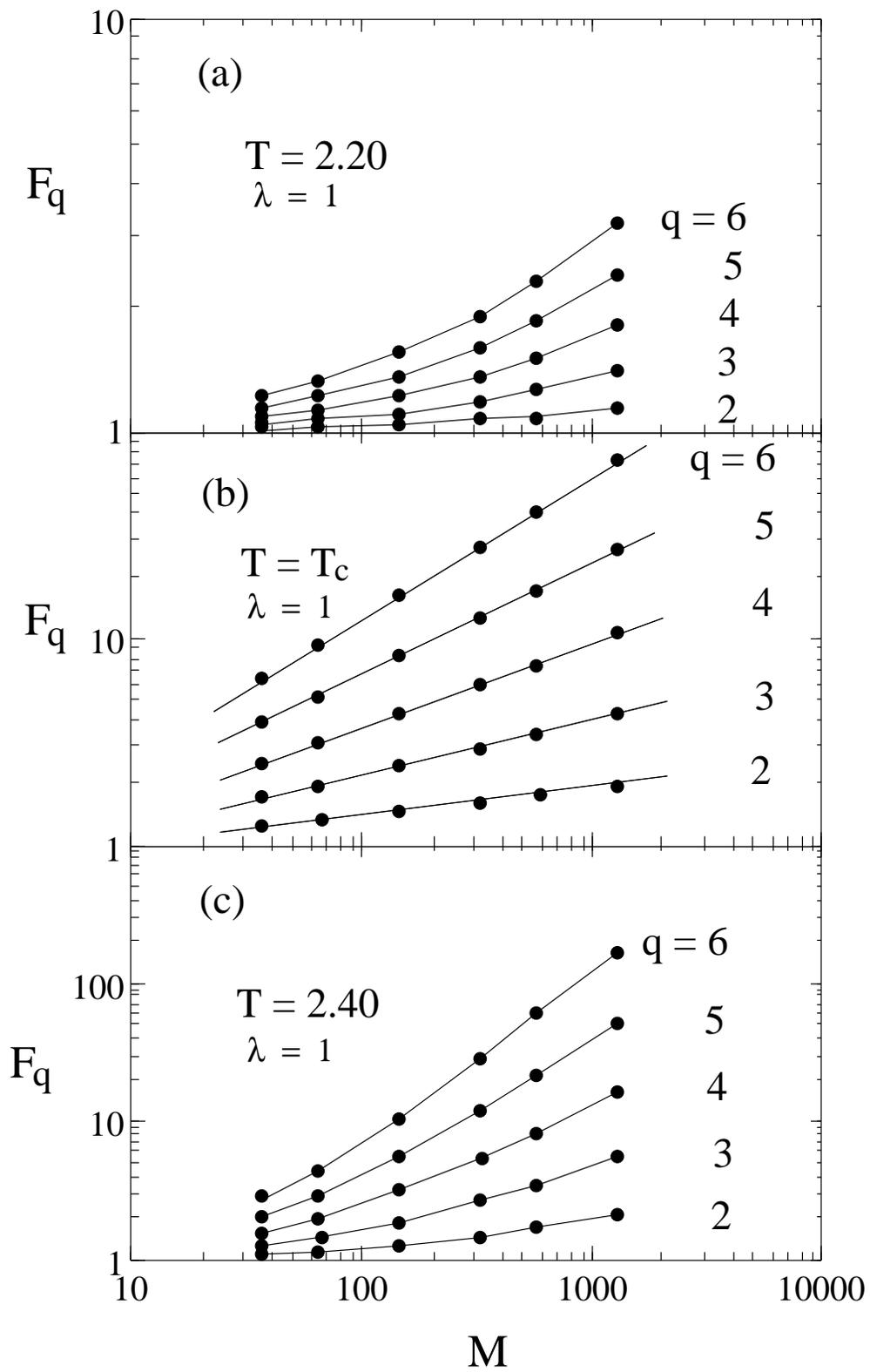
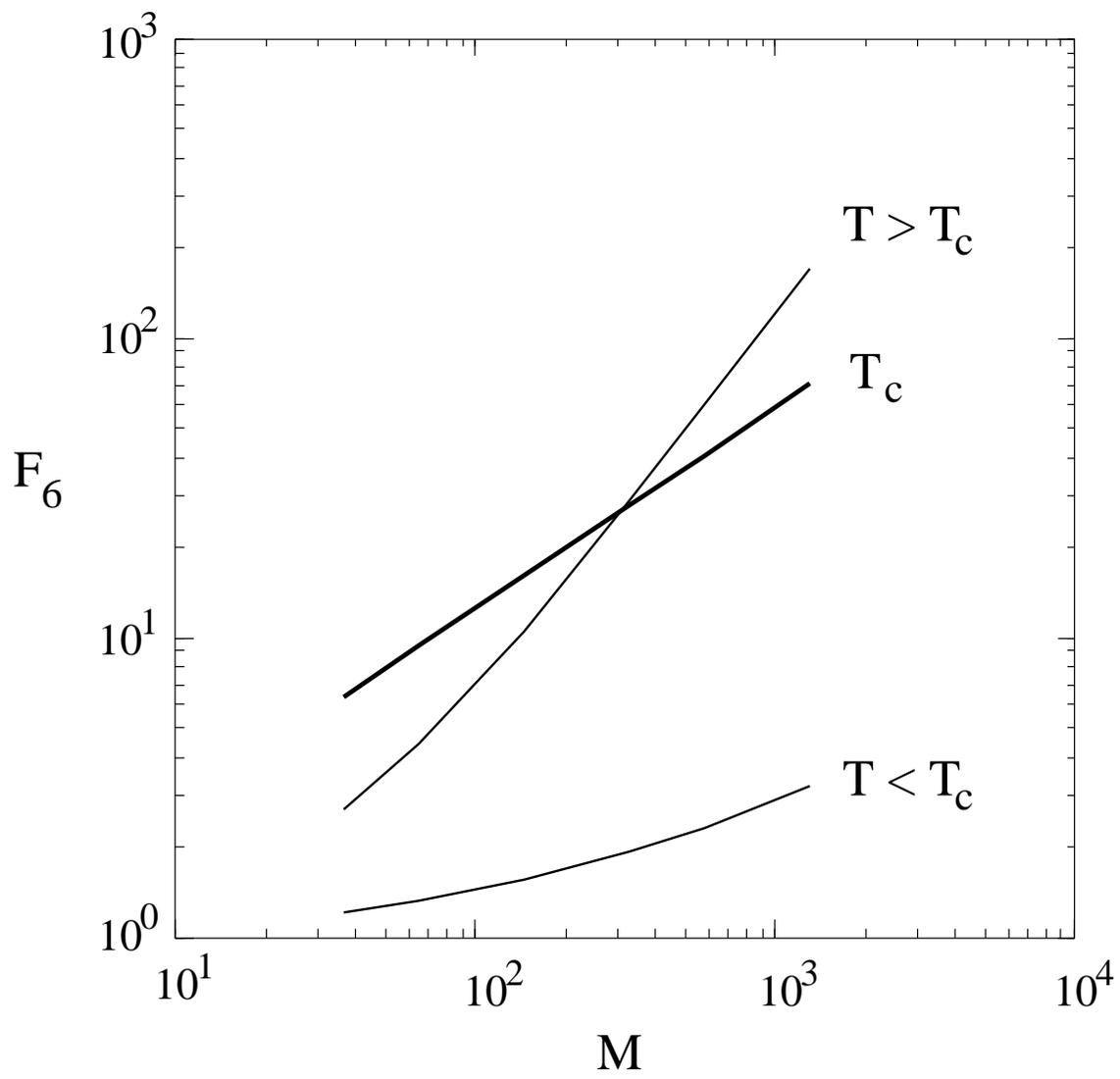
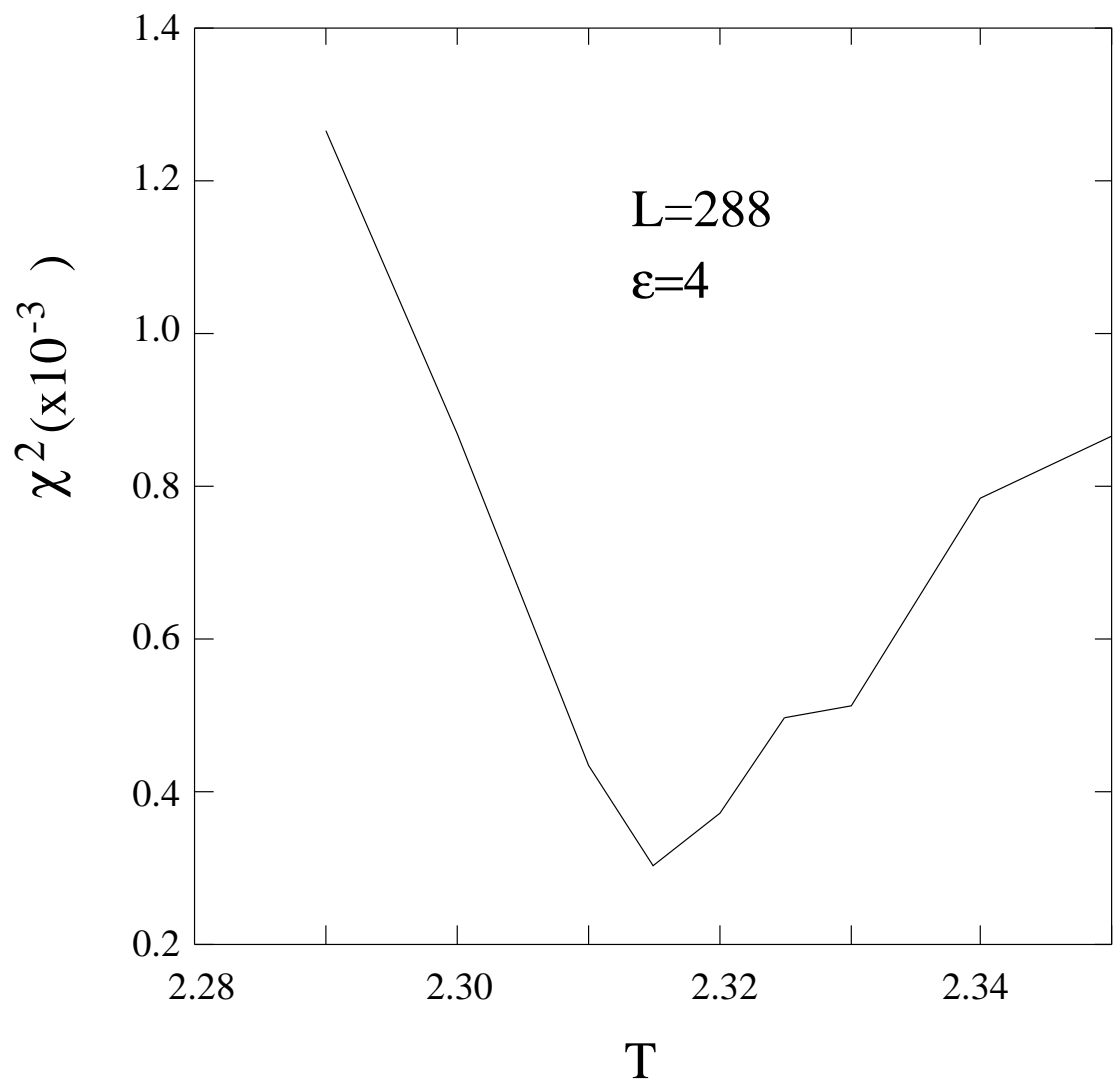
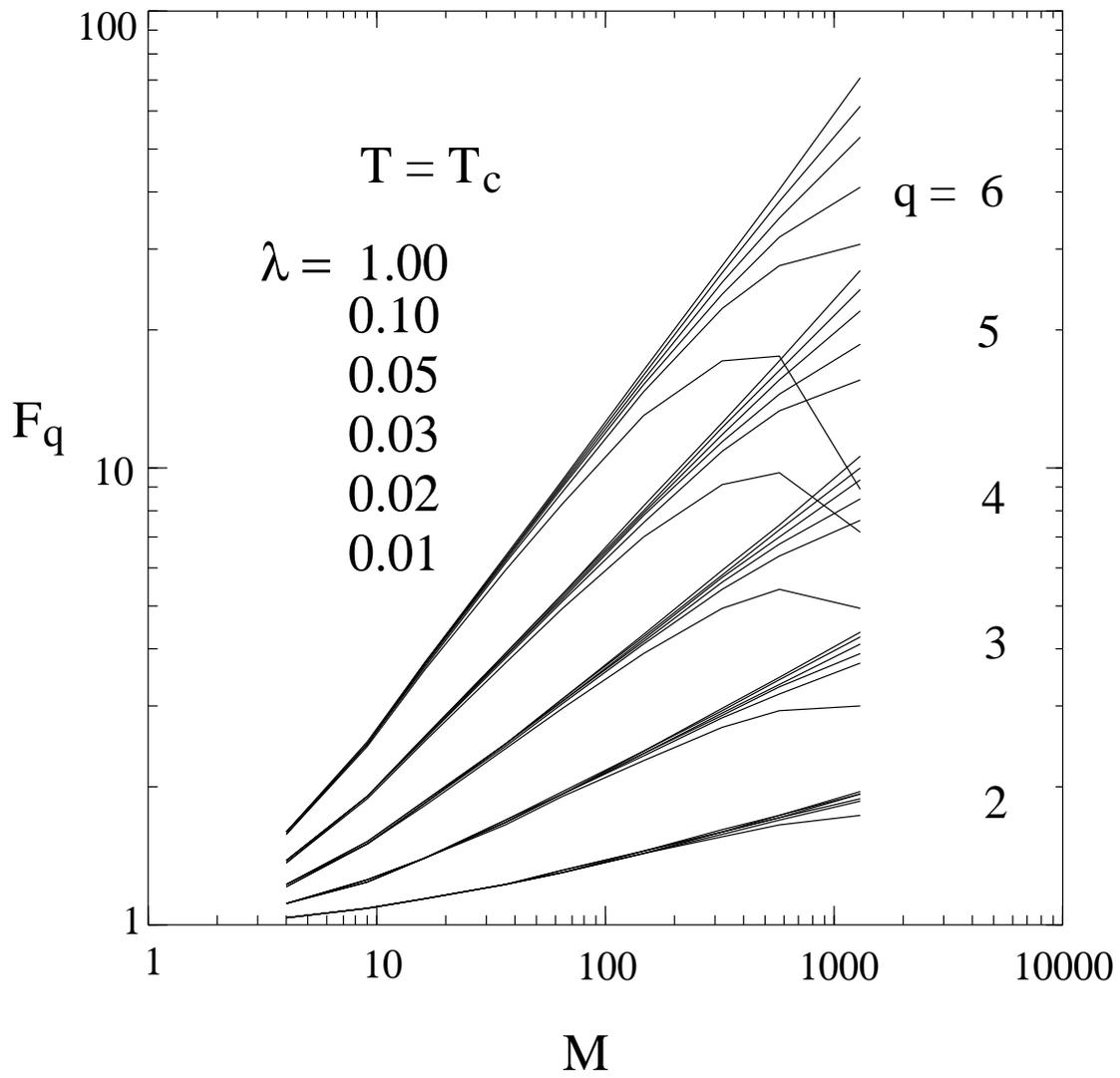
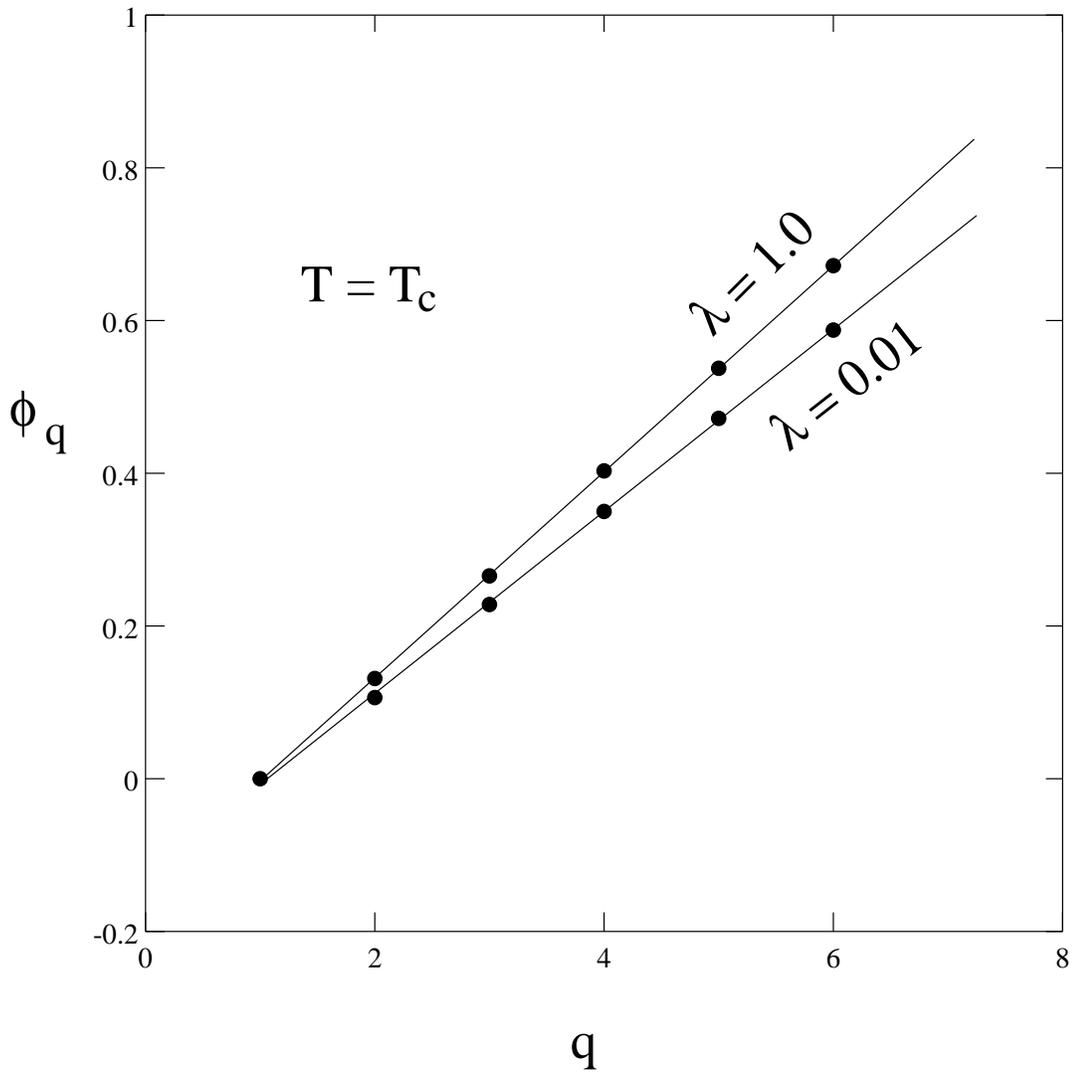
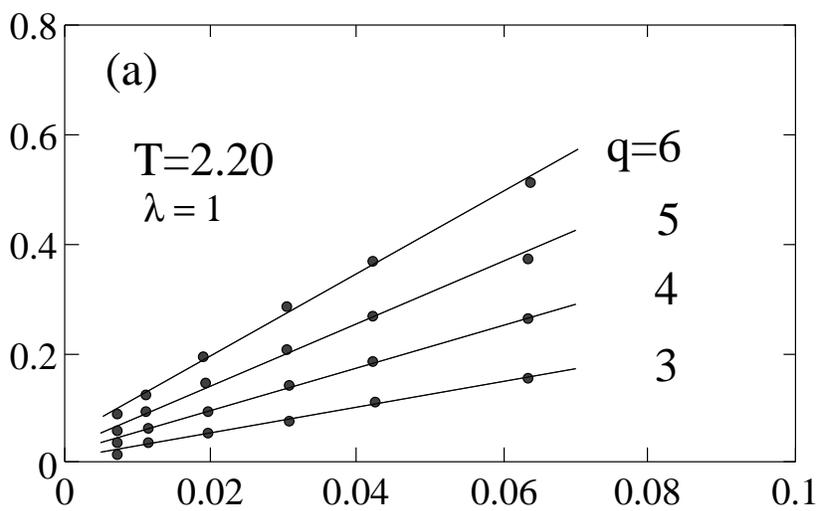
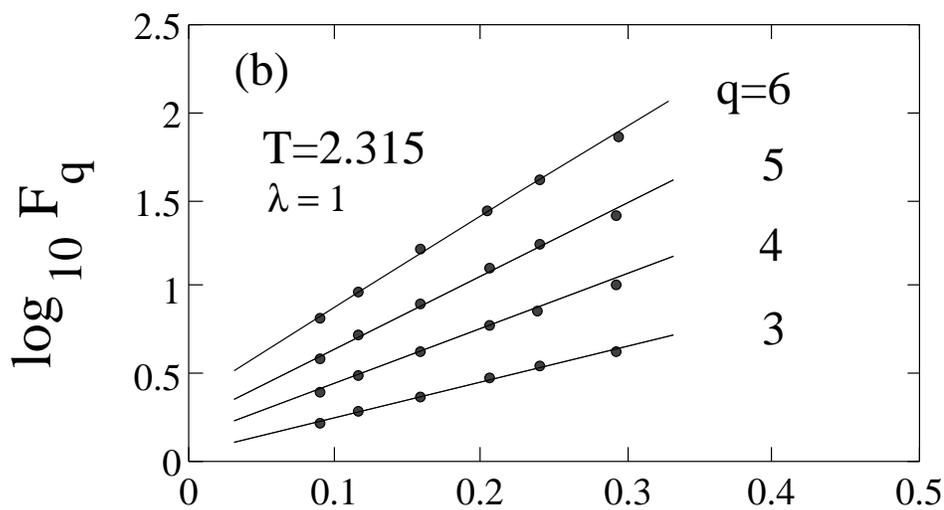
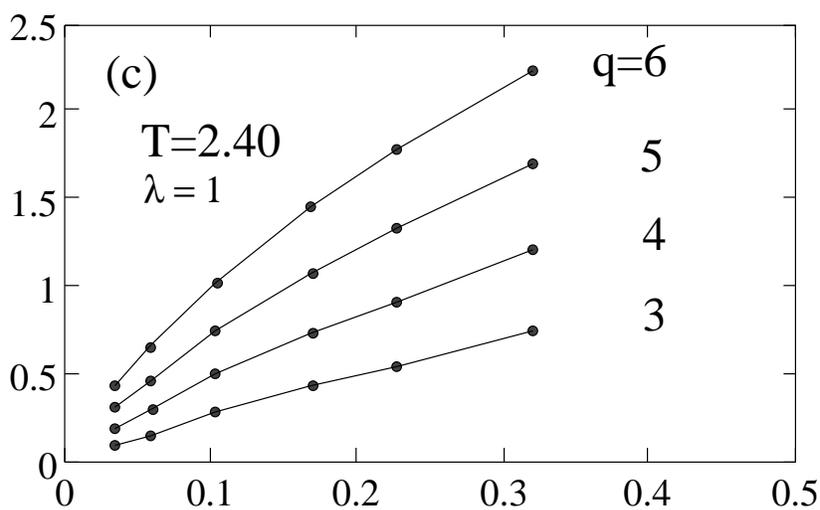
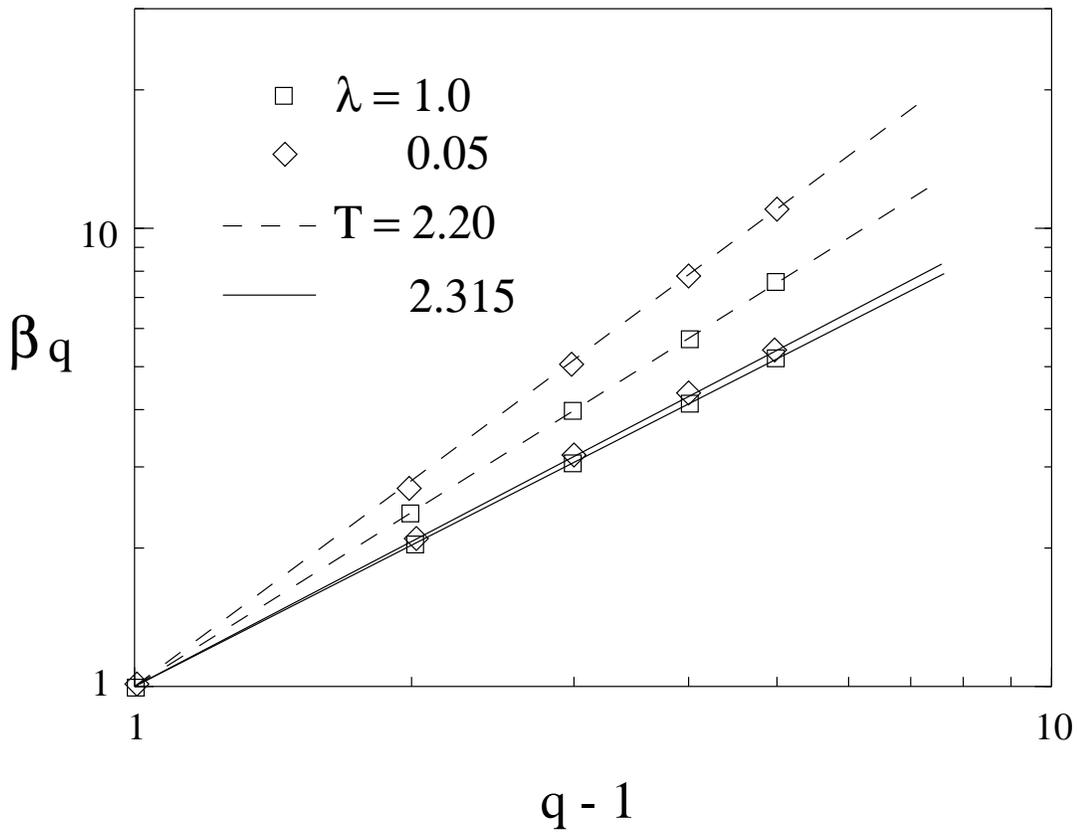
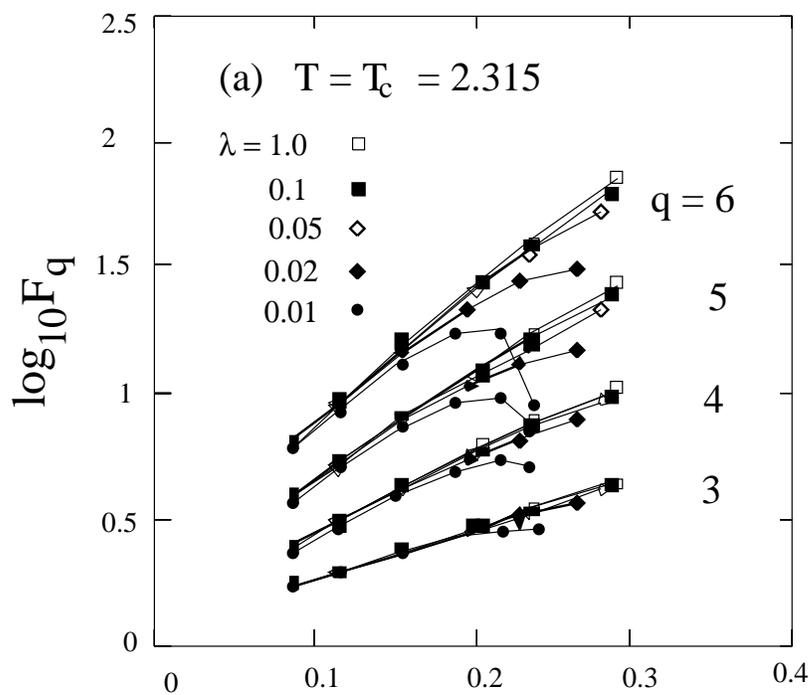
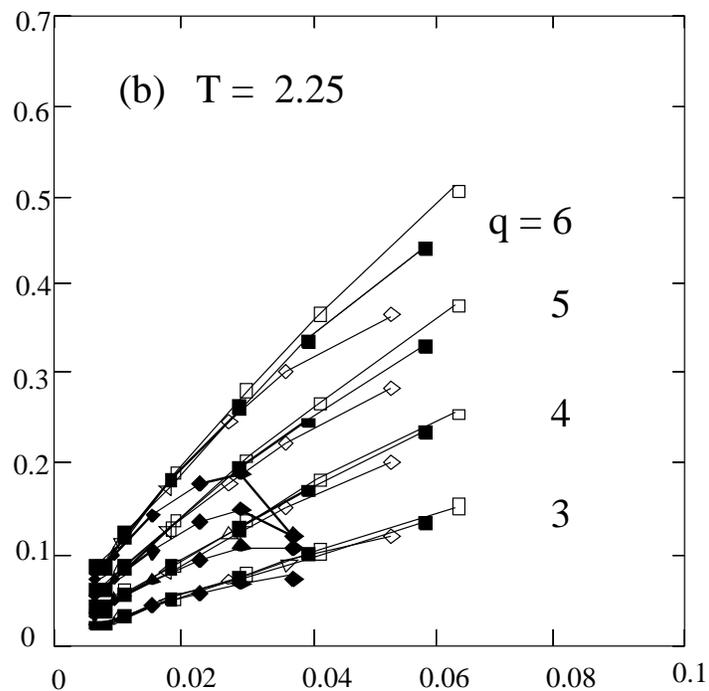
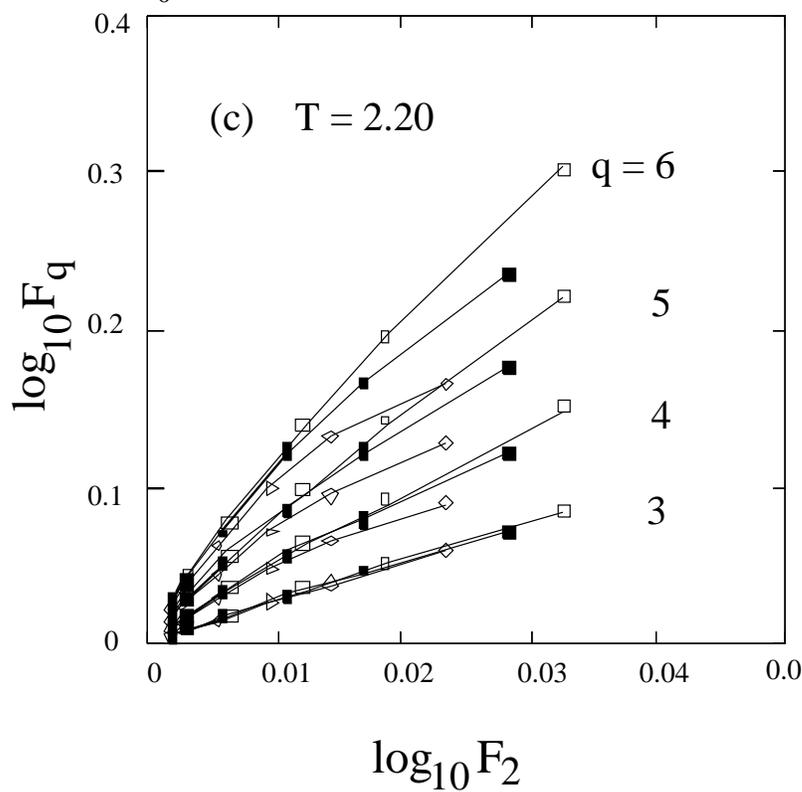
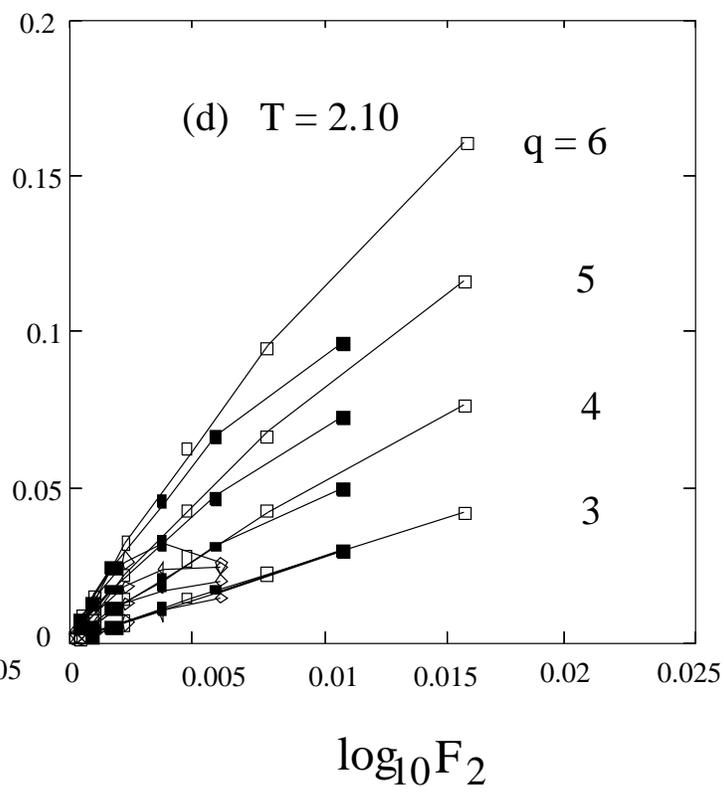
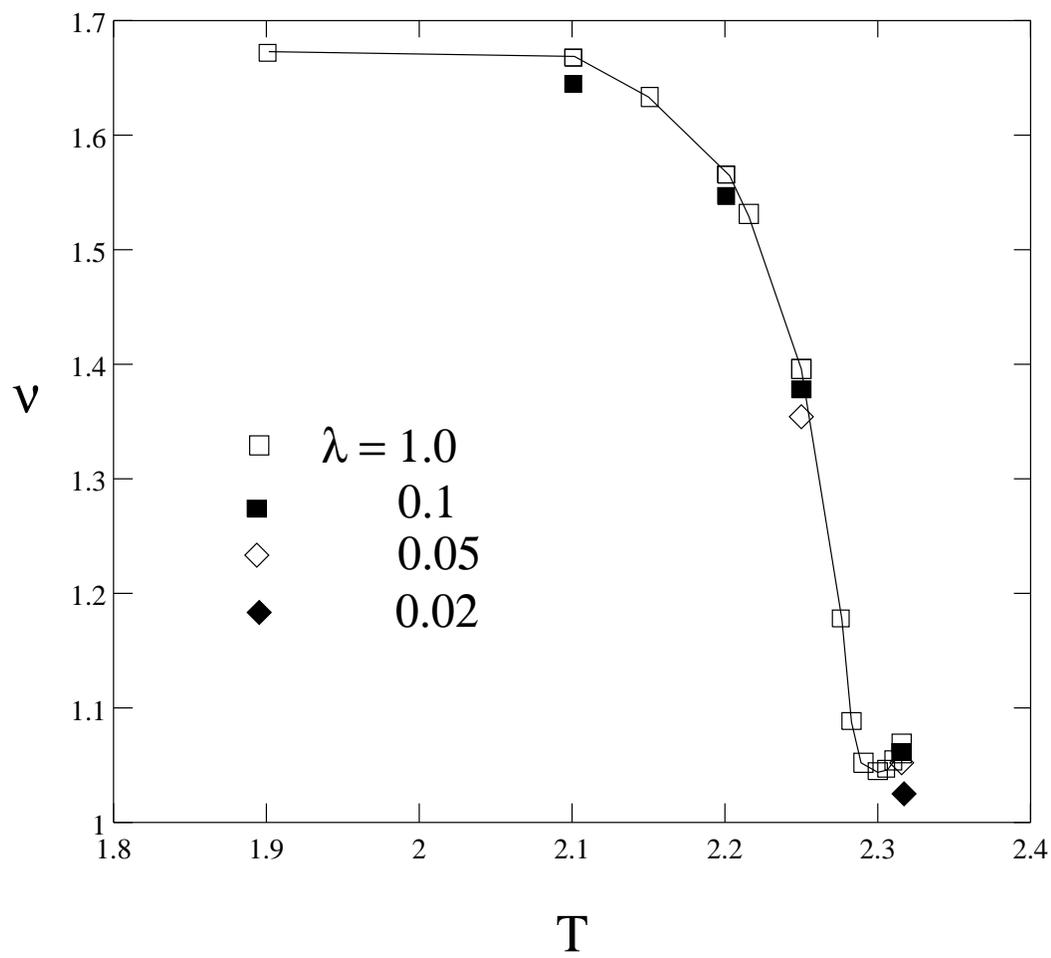
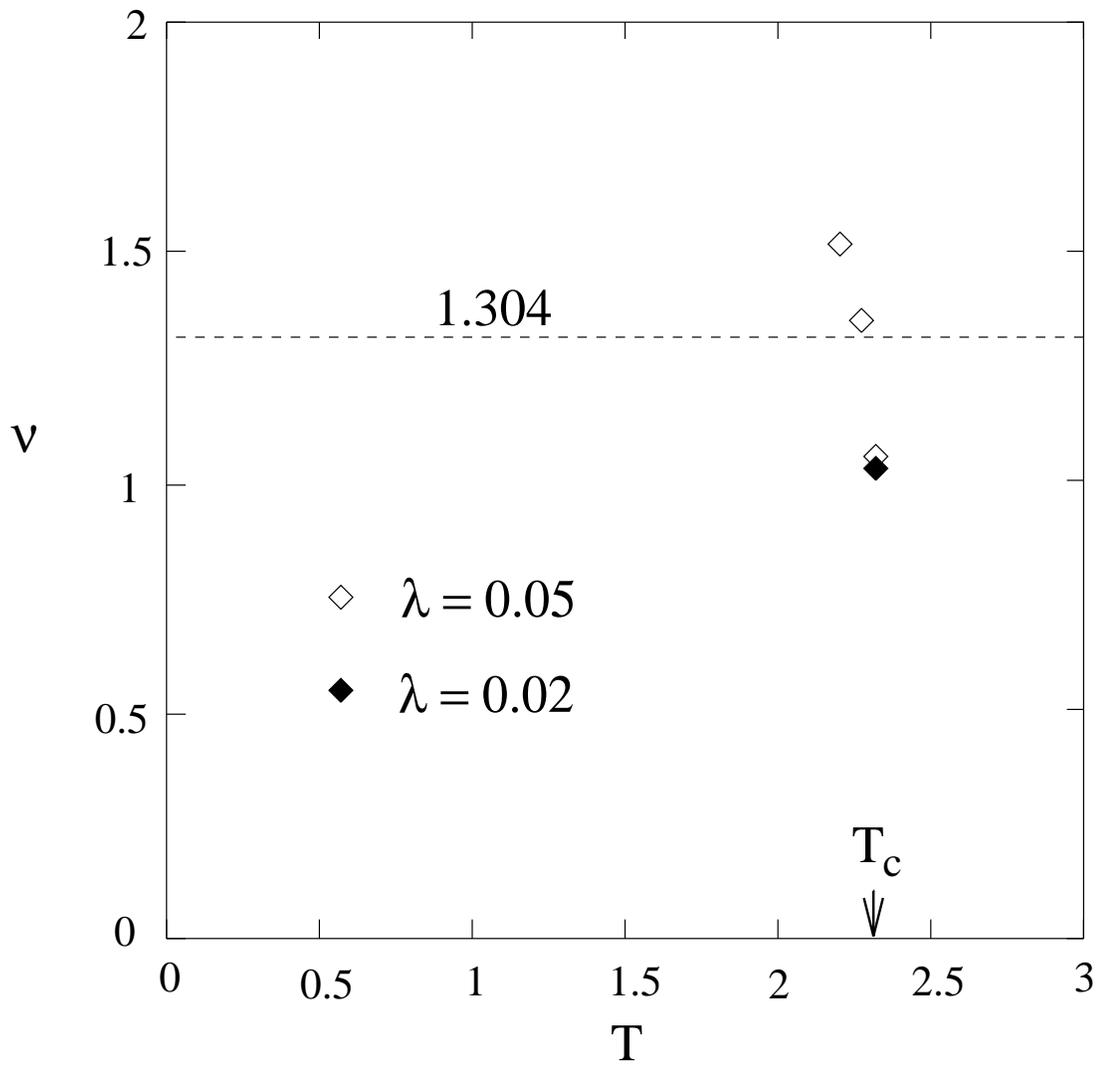
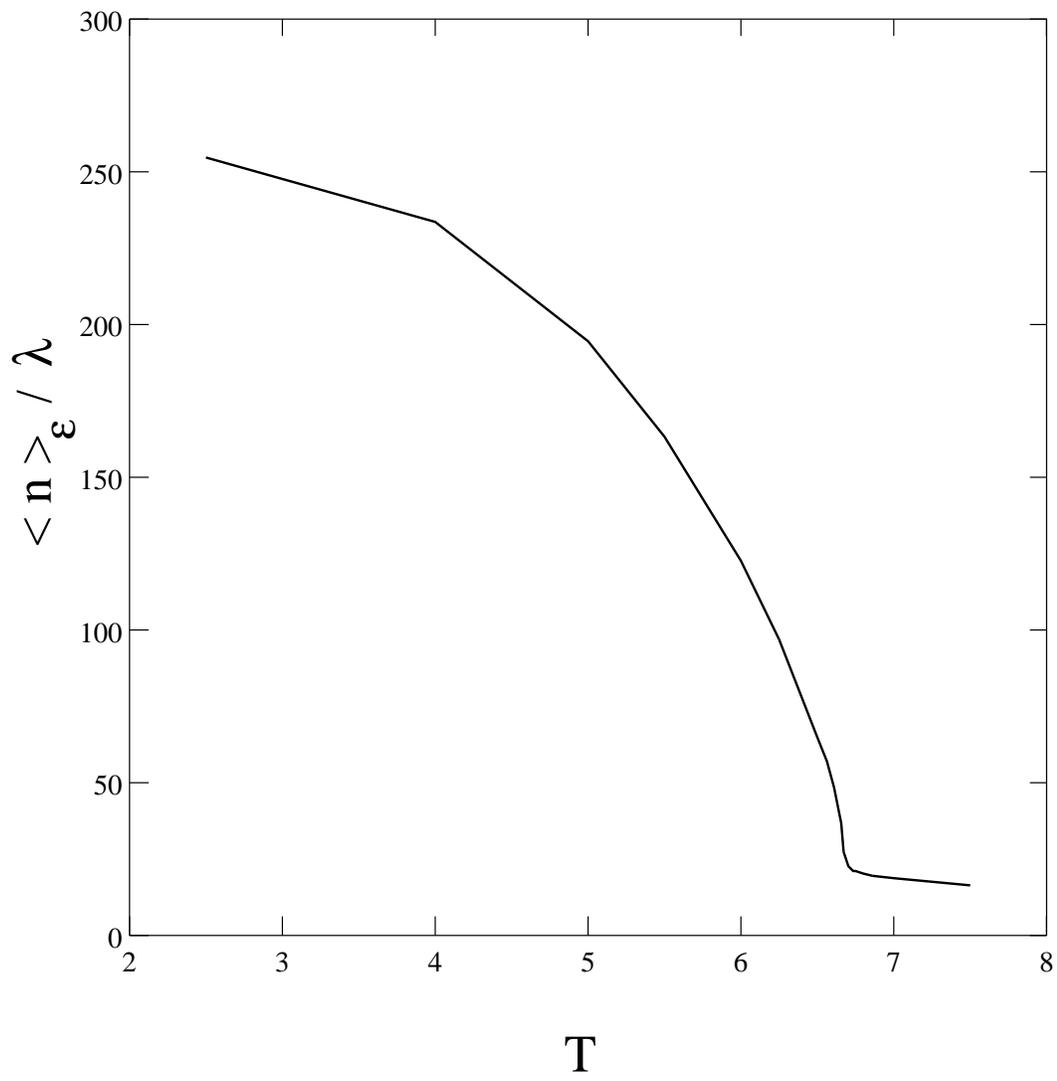
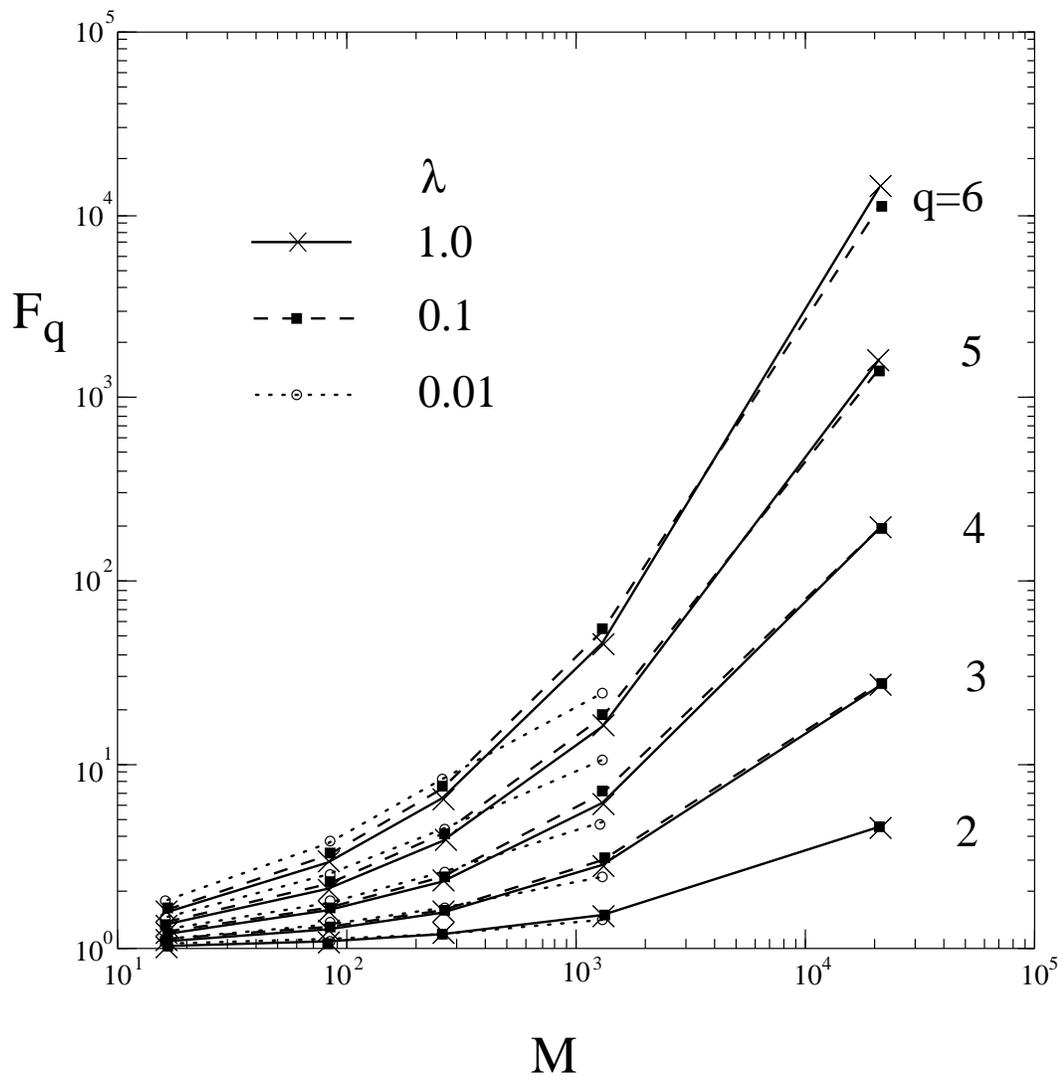
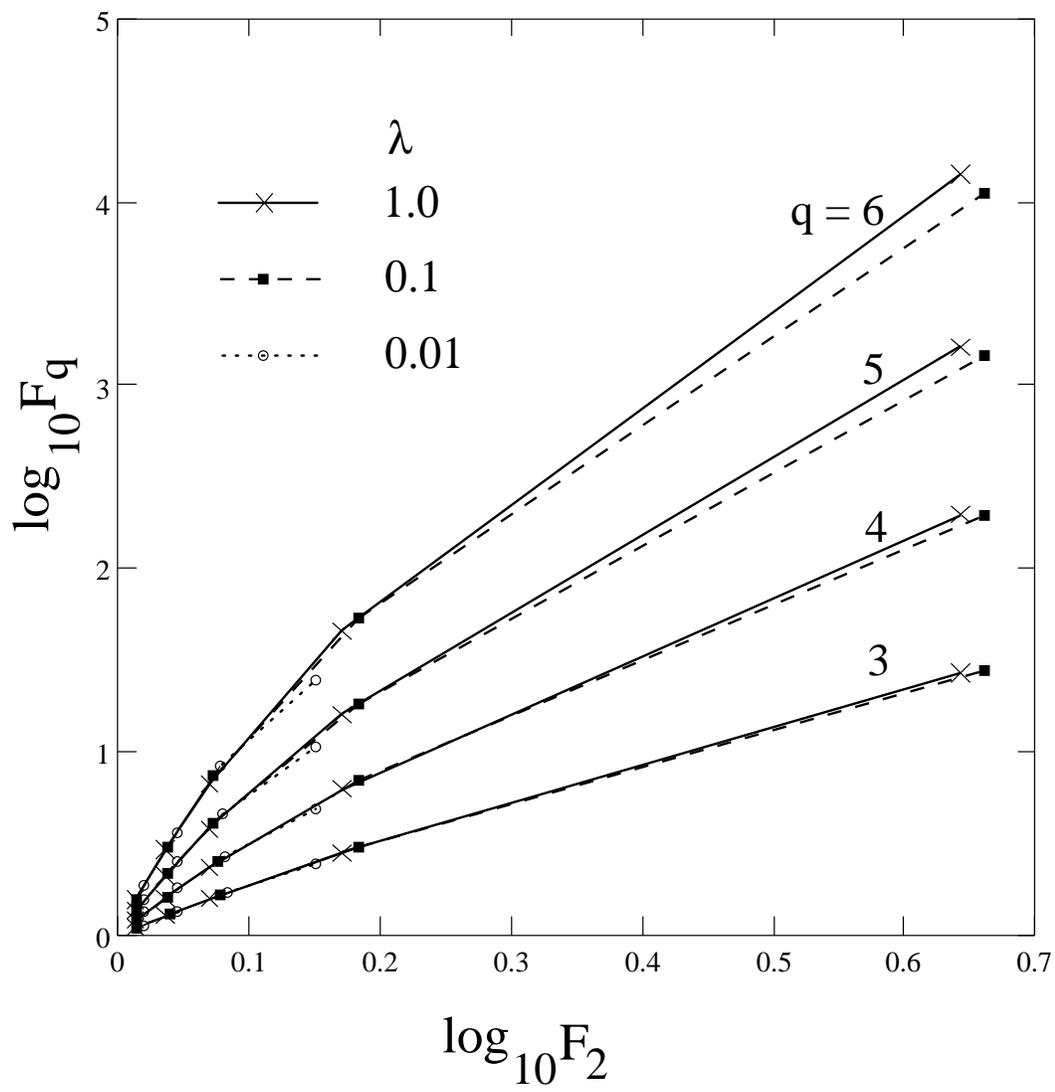
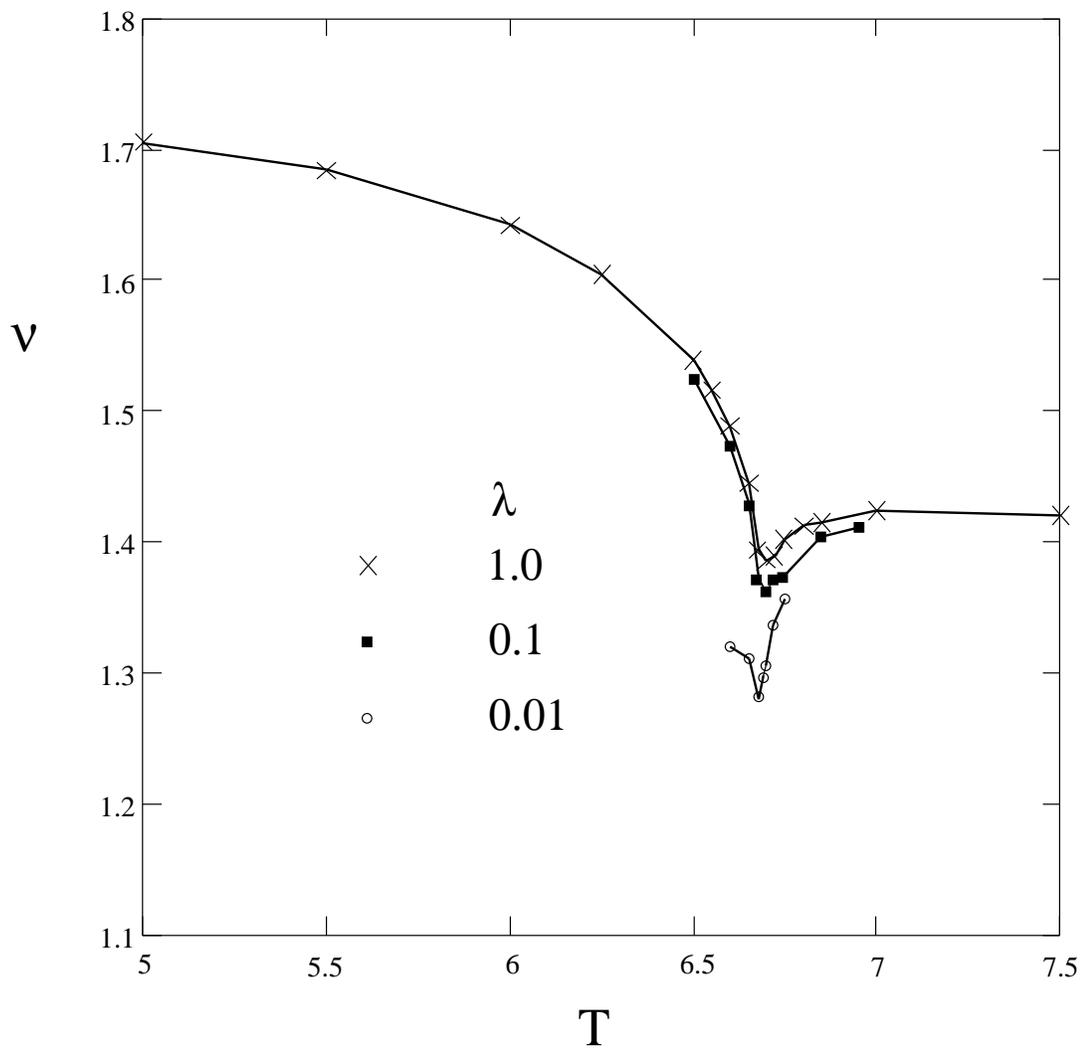
\end{document}